\documentclass[seceq]{ptptex}

\usepackage[dvips]{graphicx}
\usepackage{epsfig}

\newcommand{\lsim}{ \mathop{}_{\textstyle \sim}^{\textstyle <} }
\def\tA1{\tilde{A_1}}
\def\tA2{\tilde{A_2}}

\def\al{\alpha}

\def\lm{\lambda}
\def\Lm{\Lambda}

\title{
Localized Vector Multiplet on a Wall 
}

\author{
Nobuhito \textsc{Maru}$^{1,}$\footnote{e-mail: 
maru@postman.riken.go.jp, Special Postdoctoral Researcher} 
and Norisuke \textsc{Sakai}$^{2,}$\footnote{e-mail: 
nsakai@th.phys.titech.ac.jp
}
}

\inst{
$^{1}$Theoretical Physics Laboratory, RIKEN \\
Wako 351-0198, Japan \\
$^{2}$Department of Physics, Tokyo Institute of Technology \\
Tokyo 152-8551, Japan  
}

\abst{
The localization of vector multiplets is examined 
using the 
${\cal N}=1$ supersymmetric $U(1)$ gauge theory 
with 
the Fayet-Iliopoulos term coupled to charged chiral 
multiplets in four dimensions. 
The vector field becomes localized on 
a BPS wall connecting two different vacua 
that break the gauge symmetry. 
The vacuum expectation values of charged fields 
vanish (approximately) around the center of the wall, 
causing the Higgs mechanism to be ineffective. 
The mass of the localized vector multiplet is found to be 
the inverse width of the wall. 
The model gives an explicit example of this general phenomenon. 
A five-dimensional version of the model can also be 
constructed if we abandon supersymmetry. 
}

\begin{document}

\maketitle

\section{Introduction}
In the brane-world scenario, 
\cite{LED}\tocite{HW} 
our four-dimensional world 
is realized on topological defects such as 
domain walls. 
To make a model with extra dimensions viable, it is 
necessary to be able to 
confine the particles in the standard model 
on topological defects. 
The localization of particles on topological defects 
has been studied extensively \cite{Rubakov:2001kp}. 
Massless chiral fermions can be localized on a domain 
wall. \cite{Jackiw:1975fn}
\tocite{Akama:jy} 
Massless scalars and spinors can be obtained as 
Nambu-Goldstone particles associated with 
spontaneously broken continuous global symmetries. 
Massless gravitons have also been obtained in warped 
metric models \cite{RS}. 
Although it is difficult to 
obtain massless or nearly massless vector bosons 
in field theories, 
massless vector bosons are localized on D-branes 
in string theory. 
There have been some proposals for vector boson 
localization in field theories. 
One of them uses confined vector 
bosons that are deconfined near a topological defect 
\cite{Dvali:1996xe}. 
This mechanism employs nonperturbative effects that are 
somewhat difficult to realize explicitly, and may be difficult 
to implement, especially in higher dimensions. 
Another series of interesting proposals has been made using 
gravitational interactions 
in a vortex background in a warped six dimensional system. 
\cite{Gherghetta:2000qi}\tocite{Giovannini:2002mk} 
Considered naively, 
the localization of vector bosons with 
minimal kinetic terms is impossible 
in five dimensions, 
even in the warped case \cite{Pomarol}, 
because the theory is scale invariant. 
Some extensions of vector boson localization 
in warped five dimensions have been studied. 
\cite{Kehagias}\tocite{GTU} 
These gravitational mechanisms are interesting, 
but it is perhaps more desirable to explore 
mechanisms 
that are valid even without gravitational interactions.

It has been useful to implement supersymmetry (SUSY) 
in the construction of unified models beyond the standard model 
\cite{DGSW}. 
SUSY also helps us to obtain topological defects as 
states preserving a part of SUSY. 
These are called BPS states, \cite{BPS,WittenOlive} 
and they are guaranteed to be 
minimal energy states as long as the boundary 
condition is maintained. 
SUSY theories are also useful for implementing the localization 
of particles. 
For instance, there is 
a recent proposal for gauge multiplet localization 
in terms of an ${\cal N}=2$ SUSY 
field theory in four dimensions 
with vector and hypermultiplets. 
 \cite{ShifmanYung}
The goal of that study is to construct 
a more concrete perturbative realization of 
the proposal given in Ref.9).

The purpose of this paper is to examine a 
concrete model of a possible 
localization mechanism of vector multiplets using 
a gauge theory that is spontaneously broken 
except near the domain wall. 
If the gauge symmetry is restored inside the wall, 
the vector multiplet will only freely propagate 
inside the wall. 
To implement this feature, we use a toy model 
with ${\cal N}=1$ SUSY $U(1)$ gauge theory 
in four dimensions with the 
Fayet-Iliopoulos term. 
Two charged chiral multiplets are introduced 
with a superpotential that admits two discrete SUSY 
vacua. 
The $U(1)$ gauge symmetry is broken in these vacua. 
We construct a BPS domain wall solution interpolating 
between these vacua. 
In the middle of the wall, the vacuum expectation values 
of charged scalar fields vanish (or nearly vanish), and the 
Higgs mechanism becomes ineffective locally. 
Therefore, we obtain a 
vector boson and its superpartner (gaugino) 
localized on the wall. 
The gaugino is forced to be localized also in our 
model because of the partial preservation of SUSY. 
The gaugino becomes massive through a Yukawa-type interaction 
between the charged scalar and the chargino 
when the symmetry is broken. 
In the middle of the wall, the charged scalar fields 
vanish (approximately), and the gaugino-chargino mixing is 
lost locally 
near the center of the wall, similarly to the Higgs 
mechanism for the vector boson lost there. 
As a result, a localized 
gaugino is obtained. 
In this respect, the gaugino localization in this mechanism 
has a similarity to the mechanism of the chiral fermion 
localization on a wall \cite{Jackiw:1975fn}. 

In our concrete model, we force the charged scalars to vanish 
(approximately) in the center of the wall. 
As a result, the mass of the vector multiplet 
turns out to be small, but it is of the same order 
of magnitude as 
the inverse width of the wall. 
Because charged fields condense outside the wall, 
the superconducting bulk can absorb any electric flux 
originating from test charged particles placed on the wall. 
Therefore the electric charges of the test particles are screened, 
resulting in a massive photon. 
This is the reason why the mass of the vector multiplet is given by 
the inverse width of the wall. 
Our explicit model gives a concrete example of this general 
qualitative behavior \cite{Dvali:1996xe, LED}. 
If we wish to obtain a model of vector multiplet 
localization in five dimensions, we can take 
the bosonic part of our model without assuming 
SUSY, and then promote the theory to five dimensions. 
The same mechanism of vector boson localization 
certainly can be effective in this 
five-dimensional theory, but the cost is that we must 
abandon SUSY. 

More recently, a massless gauge multiplet localized on 
a wall in five dimensions has been obtained by introducing 
tensor multiplets \cite{IOS}.  

In sect.\ref{sc:model}, our model is introduced, 
and the BPS equation for the wall is solved. 
In sect.\ref{sc:mode-equation}, mode equations 
for vector bosons are defined, 
and the masses and the mode functions are obtained. 
In sect.\ref{sc:small-kappa-sol}, the BPS wall solution is examined 
in the limit of small Fayet-Iliopoulos parameter. 
Some useful details regarding the method of solving the BPS equation 
are given in the appendix. 

\section{Model and BPS wall solution}
\label{sc:model}

To obtain a model of a vector boson localized on a wall, 
we consider an ${\cal N}=1$ SUSY $U(1)$ vector 
multiplet $V$. 
We wish to have at least two discrete SUSY vacua 
that break the gauge symmetry. 
We introduce the 
 chiral scalar fields $\Phi_1$ and $\Phi_2$ 
 with unit positive and negative charge respectively, 
to avoid an anomaly. 
 We denote their scalar components as
 $A_1$ and $A_2$. 
To  form a nontrivial wall solution, we introduce 
a superpotential $P$ as a function of the product 
$\Phi_1 \Phi_2$. 
It is desirable to arrange the two SUSY vacua to have real 
field values of opposite sign, so that the charged 
field vanishes ($A_1=0, A_2=0$) in the middle of the wall 
when interpolating between two vacua with a real field 
configuration. 
Because $P$ is a function of  $\Phi_1\Phi_2$, 
the SUSY vacuum condition from the stationarity of the 
superpotential (the $F$-flatness condition) is always 
satisfied if both charged fields vanish 
 simultaneously, {\rm i.e.} 
\begin{equation}
0=-F_1^*={\partial P \over \partial A_1}=
A_2{\partial P \over \partial (A_1A_2)}, 
\quad 
0=-F_2^*={\partial P \over \partial A_2}=
A_1{\partial P \over \partial (A_1A_2)} . 
\end{equation}
The BPS solutions connect different SUSY vacua, but they cannot 
pass through a SUSY vacuum in the middle. 
Therefore we should avoid the situation in which vanishing values 
of charged fields become a SUSY vacuum if we want to connect 
opposite sign vacua through real field configurations. 
This is achieved by introducing the Fayet-Iliopoulos 
term with coefficient $\kappa$ 
for the $U(1)$ gauge field. 

To obtain nonvanishing vacuum expectation values 
for the charged field in the SUSY vacua, 
we choose $P$ to be cubic in $\Phi_1\Phi_2$ with a 
dimensionless coupling $g$ and a coupling $\Lambda$ 
of unit mass dimension. 
The Lagrangian we consider is given by\footnote{
We follow the 
convention of Ref.26). 
} 
\begin{eqnarray}
{\cal L} &=& \frac{1}{4}W^\al W_\al |_{\theta^2} 
+ \frac{1}{4}\bar{W}_{\dot\alpha} 
\bar{W}^{\dot\alpha} |_{\bar{\theta}^2} + 
\left[ 2 \kappa V 
+ \Phi_1^{\dag} e^{eV} \Phi_1 + \Phi_2^{\dag} e^{-eV} \Phi_2 \right]
_{\theta^2 \bar{\theta}^2} \nonumber \\
&&+ (P(\Phi_1, \Phi_2)|_{\theta^2} + {\rm h.c.})), 
\end{eqnarray}
with
\begin{eqnarray}
P &=& \frac{g}{\Lm^3} \Phi_1 \Phi_2 \left[ \Lm^4 -\frac{1}{3} 
(\Phi_1 \Phi_2)^2 \right]. 
\end{eqnarray}
The SUSY vacua are determined by the F- and D-flatness conditions: 
\begin{eqnarray}
\label{fflat1}
\frac{\partial P}{\partial A_1} &=& \frac{g}{\Lm^3} A_2 
\left[ \Lm^4 - (A_1 A_2)^2 \right] = 0, 
\\
\label{fflat2}
\frac{\partial P}{\partial A_2} &=& \frac{g}{\Lm^3} A_1 
\left[ \Lm^4 - (A_1 A_2)^2 \right] = 0, 
\\
\label{dflat}
-D &=& \frac{e}{2}(|A_1|^2 - |A_2|^2) + \kappa = 0. 
\end{eqnarray}
The F-flatness conditions (\ref{fflat1}) and (\ref{fflat2}) 
give only three possible discrete vacua, 
\begin{eqnarray}
\label{vacuum1}
&& A_1 A_2 = \Lm^2,~P = \frac{2}{3} g \Lm^3, \\
\label{vacuum2}
&& A_1 A_2 = -\Lm^2,~P = -\frac{2}{3} g \Lm^3, \\
\label{vacuum3}
&& A_1 = A_2 = 0,~P = 0. 
\end{eqnarray}
In the presence of the Fayet-Iliopoulos term 
$\kappa \ne 0$, 
the D-flatness condition does not allow 
the vacuum (\ref{vacuum3}), 
while it does permit the 
vacua (\ref{vacuum1}) and (\ref{vacuum2}) 
with the vacuum expectation value 
determined as 
\begin{eqnarray}
|A_1|^2 = -\frac{\kappa}{e} 
 + \sqrt{\left( \frac{\kappa}{e} \right)^2 + \Lm^4},
\quad |A_2|^2 = \frac{\kappa}{e} 
 + \sqrt{\left( \frac{\kappa}{e} \right)^2 + \Lm^4}. 
 \label{eq:VEV}
\end{eqnarray}
The phase $\alpha$ of the charged fields is an 
unphysical gauge degree of freedom: 
\begin{eqnarray}
A_1 = |A_1| e^{i\al}, \quad A_2 = \pm \frac{\Lm^2}{A_1} e^{-i\al}. 
\end{eqnarray}
Here, the upper sign corresponds to the vacuum 
(\ref{vacuum1}) and the lower to (\ref{vacuum2}). 
The bosonic part of the Lagrangian is given by 
\begin{eqnarray}
{\cal L}_{{\rm boson}} &=& -\frac{1}{4} v_{mn} v^{mn} 
- {\cal D}_m A_1^* {\cal D}^m A_1 
- {\cal D}_m A_2^* {\cal D}^m A_2 
-\frac{1}{2} \left( \frac{e}{2}(|A_1|^2 - |A_2|^2) 
+ \kappa \right)^2 
\nonumber \\
&& -\left| \frac{g}{\Lm^3} (\Lm^4 -(A_1 A_2)^2) \right|^2 
(|A_1|^2 + |A_2|^2), 
\label{eq:boson_Lag}
\end{eqnarray}
\begin{equation}
{\cal D}_m A_1=\partial_m  A_1+ {i \over 2}ev_m A_1, 
\qquad 
{\cal D}_m A_2=\partial_m  A_2- {i \over 2}ev_m A_2. 
\end{equation}
After the spontaneous symmetry breaking, 
all the particles become massive.

We assume 3-dimensional Lorentz invariance and 
 take $x^2=y$ as the extra coordinate. 
The BPS equations for the chiral scalar fields read 
\cite{CQR,OINS}
\begin{eqnarray}
\label{BPS1}
\frac{d A_1}{d y} 
&=& \frac{\partial P^*}{\partial A_1^*} 
= \frac{g}{\Lm^3} A_2^* [\Lm^4 -(A_1 A_2)^2], \\
\label{BPS2}
\frac{d A_2}{d y} 
&=& \frac{\partial P^*}{\partial A_2^*} 
= \frac{g}{\Lm^3} A_1^* [\Lm^4 -(A_1 A_2)^2]. 
\end{eqnarray}
Because of the three-dimensional Lorentz invariance, 
the BPS equation for the vector multiplet becomes trivial: 
\begin{equation}
0=-D = \frac{e}{2}(|A_1|^2 - |A_2|^2) + \kappa, 
\quad v_{mn}=0. 
\label{Be2vector}
\end{equation}
Let us take the boundary conditions at $y \to \pm\infty$ 
to be real. 
Then the BPS equations dictate that the 
solutions must be real, {\rm i.e.} $A_i^*=A_i$. 
Moreover, we obtain 
\begin{eqnarray}
\frac{dA_1^2}{d y} &=& \frac{dA_2^2}{dy} = 2A_1 \frac{dA_1}{dy} 
= \frac{g}{\Lm^3}2 A_1 A_2 [\Lm^4 -(A_1 A_2)^2], 
\end{eqnarray}
which shows that 
$A_1^2 - A_2^2$ is independent of $y$, and hence 
is given by the vacuum value $ -\frac{2}{e} \kappa$. 
Therefore the BPS equation for the vector multiplet 
(\ref{Be2vector}) is 
automatically satisfied. 
Thus, our BPS equation reduces to the integrable 
equation 
\begin{eqnarray}
\frac{dA_1}{dy} = \sqrt{A_1^2 +\frac{2}{e} \kappa} 
\left[ \Lm^4 - A_1^2 \left(A_1^2 + \frac{2}{e} \kappa \right) \right]. 
\end{eqnarray}
It is convenient to introduce the rescaled variables 
\begin{eqnarray}
\tilde{A}_1 = {A_1 \over \Lambda}, 
\qquad 
 \tilde{y}={g \Lambda} y,
\qquad 
 \tilde{\kappa}= \frac{ \kappa }{e \Lambda^2}. 
\label{eq:rescaled-variable}
\end{eqnarray}
The vacuum values (\ref{eq:VEV}) for 
these rescaled fields $\tilde{A}_1^2$ and $\tilde{A}_2^2$ 
are given by 
\begin{equation}
a 
\equiv -\tilde \kappa + \sqrt{1+\tilde \kappa^2}, 
\qquad 
b 
\equiv \tilde \kappa + \sqrt{1+\tilde \kappa^2}, 
\end{equation}
respectively. 
As shown in Appendix \ref{ap:BPSeq}, we obtain the exact solution 
with the position of the center of wall $y_0$ as 
a modulus: 
\begin{eqnarray}
&&\tilde{y} - \tilde{y}_0 = \frac{1}{a+b}
\left[ \frac{1}{2\sqrt{a(a + 2\tilde{\kappa})}} 
{\rm ln} \left(\frac{(\tilde{A} + \sqrt{a})
[\sqrt{(a + 2\tilde{\kappa})
(\tilde{A}^2 + 2\tilde{\kappa})} 
+\sqrt{a} \tilde{A} + 2\tilde{\kappa}]}
{(\sqrt{a}-\tilde{A})[\sqrt{(a + 2\tilde{\kappa})
(\tilde{A}^2 + 2\tilde{\kappa})} 
- \sqrt{a} \tilde{A} + 2\tilde{\kappa}]} \right) 
\right. \nonumber \\
&& \left. - \frac{1}{2\sqrt{-b(-b + 2\tilde{\kappa})}} 
{\rm ln} \left(\frac{(\tilde{A} + \sqrt{-b})
[\sqrt{(-b + 2\tilde{\kappa})
(\tilde{A}^2 + 2\tilde{\kappa})} 
+\sqrt{-b} \tilde{A} + 2\tilde{\kappa}]}
{(\sqrt{-b}-\tilde{A})[\sqrt{(-b + 2\tilde{\kappa})
(\tilde{A}^2 + 2\tilde{\kappa})} 
- \sqrt{-b} \tilde{A} + 2\tilde{\kappa}]} \right)
\right]. 
\label{eq:BPSsol}
\end{eqnarray}
We find the asymptotic behavior of this BPS solution 
(\ref{eq:BPSsol}) for 
$y \to \pm\infty$ to be 
\begin{equation}
 \tilde{A} 
\sim 
\pm \left(
 \sqrt{a} - 
e^{\mp 2(a+b)\sqrt{a(a + 2 \tilde{\kappa})}
(\tilde{y} - \tilde{y}_0)}\right),
\end{equation}
and $\tilde{A} \to 0$ 
near the center of the wall, {\rm i.e.} as 
$\tilde y \to \tilde y_0$.

\section{Mode equation of the vector boson on the wall}
\label{sc:mode-equation}

To find the mass and wave function of the 
vector multiplet, we consider the 
equation of motion of the vector boson, 
\begin{eqnarray}
&\!\!\!&0 = \frac{\partial {\cal L}}{\partial v^n} 
- \partial^m \frac{\partial {\cal L}}{\partial \partial^m v^n} 
\\
&\!\!\!=& \partial_m \partial^m v_n 
-\partial_n \partial^m v_m 
+ \frac{i}{2} e (A_1^* \partial_n A_1 
-\partial_n A_1^* A_1 
- A_2^* \partial_n A_2 + \partial_n A_2^* A_2) \\
&&-\frac{e^2}{2} v_n (|A_1|^2 + |A_2|^2). \nonumber 
\end{eqnarray}
In order to make the Higgs mechanism explicit, 
it is better 
 to use the unitary gauge and absorb 
 $\partial_n (A_i-A_i^{*})~(i=1,2)$ 
 into the longitudinal component of 
 $v_n$, at least near the limits $y \to \pm \infty$. 
Let us consider the following nonlinear field 
redefinition to absorb the Nambu-Goldstone boson 
 into massive vector: 
\begin{eqnarray}
A_1(x,y) =  A_{1R}(x,y)e^{i \xi(x,y)}, \qquad
A_2(x,y) =  A_{2R}(x,y)e^{-i \xi(x,y)}. 
\end{eqnarray}
Here, $A_{1(2)R}$ can be taken to be real. 
After these redefinitions, 
 the equation of motion for the vector $v_n$ becomes
\begin{eqnarray}
0 &=& \partial_m \partial^m v_n -\partial_n \partial^m v_m 
-e\left[A_{1R}^2 +  A_{2R}^2\right]\partial_n \xi
-\frac{e^2}{2} \left[ A_{1R}^2 +  A_{2R}^2\right]v_n. 
\end{eqnarray}
Note that the above equation of motion is invariant 
 under the gauge transformation
\begin{eqnarray}
A_1 \to A_1e^{i \lm},\quad A_2 \to A_2 e^{-i \lm}, \quad 
v_n \to v_n - \frac{2}{e} \partial_n \lm. 
\end{eqnarray}
By the gauge transformation $\lm$, 
 we can eliminate $\xi$, thereby arriving at the unitary gauge. 
Because the terms linear in $A_{iR}$ disappear, we no longer 
need to consider fluctuations of scalar fields in the 
linearized equations of motion for the vector fields. 

We denote the coordinates in the 
three-dimensional world volume on the wall 
by the Greek indices $\mu, \nu = 0,1,3$, 
as opposed to the Roman indices $m, n = 0,1,2,3 $ of 
the fundamental theory in four dimensions. 
We obtain the linearized equations of motion  
\begin{eqnarray}
0 
&=& \partial^m \partial_m v_\nu 
-\partial_\nu \partial^m v_m 
-\frac{e^2}{2}
\left[(A_1^{{\rm cl}})^2 
+ (A_2^{{\rm cl}})^2\right]v_\nu
\nonumber \\
&=& \square v_\nu 
+  v_\nu'' 
-\partial_\nu \partial^\mu v_\mu 
-\partial_\nu v_y' 
-V(y)v_\nu 
\label{eq:lin-EOM-v}
\end{eqnarray}
for $n = \nu \ne y$, where ${}'$ denotes differentiation with respect to $y$, 
$\square \equiv \partial^\mu \partial_\mu$, and 
the potential $V(y)$ is defined as 
\begin{eqnarray}
V(y)&=&\frac{e^2}{2} \left[(A_1^{{\rm cl}})^2 
+ (A_2^{{\rm cl}})^2\right]
=
e\kappa \; {\rm cosh}\left( \frac{2}{\Lambda} 
\sqrt{\Lambda^4 + \left(\frac{\kappa}{e}\right)^2}y \right)
\nonumber \\
&=&
e^2\tilde\kappa\Lambda^2 {\rm cosh}\left( {2}{\Lambda} 
\sqrt{1 + \tilde{\kappa}^2}y \right). 
\label{eq:potential}
\end{eqnarray}
The linearized equation of motion for $n = y$ is given by 
\begin{eqnarray}
0 
&=& \partial^m \partial_m v_y -\partial_y \partial^m v_m 
-\frac{e^2}{2} \left[(A_1^{{\rm cl}})^2 
+ (A_2^{{\rm cl}})^2\right]v_y
\nonumber \\
&=& \square v_y - \partial^\mu v_\mu' 
-V(y)v_y
. 
\label{eq:lin-EOM-vy}
\end{eqnarray}

As shown in Appendix \ref{ap:vector-mode}, we find 
that there are no zero modes for the vector 
field. 
Therefore, we can decompose the vector field into 
a transverse component $\tilde v_\mu$ and a longitudinal 
component $\partial^\mu v_\mu$ as 
\begin{eqnarray}
\tilde v_\mu 
&\equiv& 
 v_\mu - {1 \over \square} 
\partial_\mu \partial^\lambda v_\lambda 
,
\label{eq:transverse-comp}
\end{eqnarray}
satisfying 
\begin{eqnarray}
\partial ^\mu \tilde v_\mu =0 .
\end{eqnarray}
Also as shown in Appendix \ref{ap:vector-mode}, the 
transverse component satisfies 
the linearized equations of motion 
\begin{eqnarray}
0 
&=& \square \tilde v_\nu 
+  \tilde  v_\nu'' 
-V(y) \tilde v_\nu. 
\label{eq:lin-EOM-trans2}
\end{eqnarray}
The longitudinal component $\partial^\mu v_\mu$ 
can be expressed in terms of 
the scalar component $v_y$, as given in 
Eq.~(\ref{eq:long-vy}), 
and satisfies an addtional linearized equation, 
(\ref{eq:lin-EOM-vy3}), which is 
decoupled from the transverse component, as shown in 
Appendix \ref{ap:vector-mode}. 

Expanding the transverse component 
$\tilde v_\nu$ of vector field in 
the complete set of mode functions $v^{(k)}(y)$, 
\begin{eqnarray}
\tilde v_\mu(x,y) = \sum_k a_\mu^{(k)}(x) v^{(k)}(y), 
\end{eqnarray}
we obtain the fields $a_\mu^{(k)}(x)$ 
in the effective three-dimensional theory as 
expansion coefficients. 
The mode functions are defined by means of 
the Hamiltonian $H$ as 
\begin{eqnarray}
H \; u^{(k)} = m_{(k)}^2 u^{(k)}, 
\qquad 
\label{eq:vec-mode-eq}
H 
\equiv 
-\partial_y^2 +V(y) ,   
\end{eqnarray}
with the potential $V(y)$ in Eq.~(\ref{eq:potential}). 
Although we have yet been unable to obtain exact solutions of 
the mode equation (\ref{eq:vec-mode-eq}), 
we can give 
lower and upper bounds on the ground state 
mass squared. 
If we expand the potential around the origin and retain 
up to quadratic order terms, we obtain a harmonic oscillator 
potential that is everywhere lower than the original 
potential 
\begin{eqnarray}
V(y) 
&\simeq & V_{\rm harmonic}(y) 
+ {\cal O}(e^2\tilde \kappa \Lambda^6 y^4), 
\\
 V_{\rm harmonic}(y) 
&\equiv & 
e^2 \tilde \kappa \Lambda^2 
\left(1+ 2\Lambda^2 
\left( 1 + \tilde {\kappa}^2 \right)y^2 \right)
\le 
V(y)
. 
\end{eqnarray}
Therefore the exact ground state eigenvalue of 
Eq.(\ref{eq:vec-mode-eq}) is bounded from below by 
the ground state eigenvalue of the harmonic oscillator 
potential, which is given by 
\begin{eqnarray}
m_{\rm harmonic}^2 &=& 
 e^2 \tilde \kappa \Lambda^2 
+ e\sqrt{2 \tilde \kappa
( 1 + \tilde{\kappa}^2) } 
\Lambda^2 
 \simeq e \sqrt{2 \tilde \kappa} \Lambda^2 ,
\end{eqnarray}
because we are interested in the case in which the 
parameter $\alpha^2$ is small, {\rm i.e.}, 
\begin{equation}
\alpha^2= 
{e \sqrt{\tilde \kappa} \over 
\sqrt{1+ \tilde{\kappa}^2 }}
\ll 1 . 
\label{eq:alpha}
\end{equation}

To obtain an upper bound on the ground state eigenvalue, 
we use a variational approach. 
Because the exact potential (\ref{eq:vec-mode-eq}) becomes 
strongly repulsive as $y$ increases, 
we should choose trial functions to be strongly 
suppressed asymptotically. 
This behavior is approximated accuately by a rigid wall 
potential.\footnote{
A harmonic oscillator potential with the 
angular frequency acting as the variational parameter 
can be another choice. 
It gives a less stringent bound of order 
$\Lambda^2/\log (1/\alpha^2)$, 
instead of $\Lambda^2/(\log (1/\alpha^2))^2$. 
} 
By defining the mass scale of the potential $\mu$ as 
\begin{eqnarray}
\mu^4 \equiv e^2 \tilde \kappa 
(1 + \tilde {\kappa}^2 )\Lambda^4
, 
\end{eqnarray}
we choose as a trial function 
the ground state 
wave function for a rigid wall potential, 
\begin{eqnarray}
V_{\rm rigid}(y)=
 \left\{
\begin{array}{cc}
0, & -\mu a< y < \mu a \\
\infty & |y|>\mu a  
\end{array}
\right. ,  
\label{eq:rigid-wall}
\end{eqnarray}
where $a$ is the 
dimensionless variational parameter for the width of 
the potential. 
As shown in Appendix \ref{ap:variation}, 
the best upper bound on the 
ground state mass squared is given by 
\begin{equation}
m_{(0)}^2 < \mu^2 \left({\pi \over 
2a
}\right)^2
, 
\label{eq:best-bound}
\end{equation}
which is 
realized when the width parameter is given by 
\begin{equation}
a \approx \alpha {\rm log} {\pi \over \alpha^2}
. 
\label{eq:width}
\end{equation}
Therefore, we find the the lowest mass squared of 
the vector boson is bounded by 
\begin{equation}
2e \sqrt{\tilde \kappa 
}\; 
\Lambda^2 < 
m_{(0)}^2 < 
\left({\pi \; \Lambda \over 
2{\rm log}\left({\pi \over 
e\sqrt{\tilde \kappa}}\right)}\right)^2
. 
\label{eq:bound-mass-squared}
\end{equation}
Because the exact potential is strongly suppressed as 
$y \rightarrow \pm \infty$, we believe 
that this upper bound may be more realistic than 
the lower bound obtained using the harmonic oscillator 
approximation. 


Due to the Higgs mechanism, the nonvanishing 
charged field $A_2^{\rm cl}$ on the wall should 
contribute a term $m^2_{\rm Higgs}$ 
in the vector boson mass squared of order 
\begin{equation}
m^2_{\rm Higgs} =(e A_2^{\rm cl})^2 =2 e \kappa \ll 
m_{(0)}^2. 
\end{equation}
We recognize that the vector boson mass is primarily due to 
the screening instead of the nonvanishing charged field 
$A_2^{\rm cl}$ on the wall.

\section{Wall solution in the limiting case $|\frac{2\kappa}{e}| \ll \Lambda^2$}
\label{sc:small-kappa-sol}

To clarify the situation of the lightest vector field, 
we examine the wall solution in the limiting case 
 $|\frac{2\kappa}{e}| \ll \Lambda^2$ in which the mass of the vector 
 field becomes small.  
The BPS equations in this limit are given in terms of the 
rescaled variables (\ref{eq:rescaled-variable}) as 
\begin{eqnarray}
\label{reduced-BPS1}
\frac{d \tilde{A}_1}{d \tilde{y}} &=& \tilde{A}_2 
(1 - \tilde{A}_1^2 \tilde{A}_2^2), \\
\label{reduced-BPS2}
\frac{d \tilde{A}_2}{d \tilde{y}} &=& \tilde{A}_1 
(1 - \tilde{A}_1^2 \tilde{A}_2^2), 
\end{eqnarray}
with the $D$-flatness constraint given by 
$\tilde{A}_1^2 - \tilde{A}_2^2 = -2 \tilde{\kappa}$. 
Multiplying Eq.~(\ref{reduced-BPS1}) by $\tilde{A}_2$ and 
Eq.~(\ref{reduced-BPS2}) by $\tilde{A}_1$ and summing them 
gives 
\begin{equation}
\frac{dX}{d \tilde{y}} 
= 2 \sqrt{X^2 + \tilde{\kappa}}(1 - X^2), 
\end{equation}
where we have defined the convenient variable 
$X \equiv \tilde{A}_1 \tilde{A}_2$.  
The solution reads 
\begin{equation}
e^{4\sqrt{1+\tilde{\kappa}^2}(\tilde{y} - \tilde{y}_0)} 
= \frac{(1+X)(\sqrt{1+\tilde{\kappa}^2} 
\sqrt{X^2+\tilde{\kappa}^2} + \tilde{\kappa}^2 + X)}
{(1-X)(\sqrt{1+\tilde{\kappa}^2} 
\sqrt{X^2+\tilde{\kappa}^2} + \tilde{\kappa}^2 - X)}. 
\end{equation}
We see immediately the reflection symmetry 
\begin{equation}
X \rightarrow -X, \qquad 
\tilde y-\tilde y_0 \rightarrow 
-(\tilde y-\tilde y_0)  
\end{equation}

If $|\tilde{\kappa}| \ll X \lsim 1$, the following 
approximations hold: 
\begin{eqnarray}
\sqrt{1 + \tilde{\kappa}^2} \sqrt{X^2+\tilde{\kappa}^2} 
+\tilde{\kappa}^2 + X
&\simeq& 
2X +\frac{\tilde{\kappa}^2 (X+1)^2}{2X} 
+{\cal O}(\tilde{\kappa}^4), \\
\sqrt{1 + \tilde{\kappa}^2} \sqrt{X^2+\tilde{\kappa}^2} 
+\tilde{\kappa}^2 - X
&\simeq& 
\frac{\tilde{\kappa}^2 (X+1)^2}{2X} 
+{\cal O}(\tilde{\kappa}^4), 
\end{eqnarray}
\begin{eqnarray}
e^{4\sqrt{1+\tilde{\kappa}^2}(\tilde{y} - \tilde{y}_0)}
\simeq \frac{1+X}{1-X} 
\frac{\frac{\tilde{\kappa}^2 (1-X)^2}{2X}}{2X} 
= \frac{\tilde{\kappa}^2}{4} \left( \frac{1}{X^2} -1 \right). 
\end{eqnarray}
By defining the length of the transition region as 
\begin{equation}
\Delta \tilde{y} \equiv \frac{1}
{4\sqrt{1 + \tilde{\kappa}^2}} 
{\rm log}\frac{4}{\tilde{\kappa}^2}, 
\label{eq:transition-region}
\end{equation}
we obtain 
\begin{equation}
X^2 \simeq 
\frac{1}{1 + e^{-4\sqrt{1+\tilde{\kappa}^2}
(\tilde{y} - \tilde{y}_0 - \Delta \tilde{y})}}.
\end{equation}

For $|X| \simeq \tilde{\kappa}$, we define 
$X=\tilde \kappa \tilde X$ 
and obtain 
\begin{eqnarray}
e^{4\sqrt{1+\tilde{\kappa}^2}(\tilde{y} - \tilde{y}_0)} 
&\simeq& \frac{\sqrt{1+\tilde{X}^2} 
+ \tilde{X}}{\sqrt{1+\tilde{X}^2} -\tilde{X}}
= (\sqrt{1+\tilde{X}^2}+\tilde{X})^2, 
\end{eqnarray}
which leads to 
\begin{equation}
\tilde{X} = 
{\rm sinh}[2\sqrt{1+\tilde{\kappa}^2}
(\tilde{y} - \tilde{y}_0)]. 
\end{equation}

Therefore, the behavior of the solution as a function of 
$\tilde y$ can most conveniently be expressed in three 
separate regions: 
\begin{eqnarray}
X \simeq \left\{
\begin{array}{c}
\tilde{\kappa}/\sqrt{\tilde{\kappa}^2 
+ 4e^{-4\sqrt{1+\tilde{\kappa}^2}(\tilde{y}-\tilde{y}_0)}} \quad 
(|\tilde{\kappa}| < X \le 1) \\
\\
\tilde{\kappa}{\rm sinh}[2\sqrt{1+\tilde{\kappa}^2}(\tilde{y} - \tilde{y}_0)] 
\quad (|X| < |\tilde{\kappa}|) \\
\\
-\tilde{\kappa}/\sqrt{\tilde{\kappa}^2 
+ 4e^{4\sqrt{1+\tilde{\kappa}^2}(\tilde{y}-\tilde{y}_0)}} \quad 
(-1 \le X < -|\tilde{\kappa}|) \\
\end{array}
\right.. 
\end{eqnarray}

\begin{figure}[htb]
\begin{center}
\includegraphics[width=6cm,height=5cm]{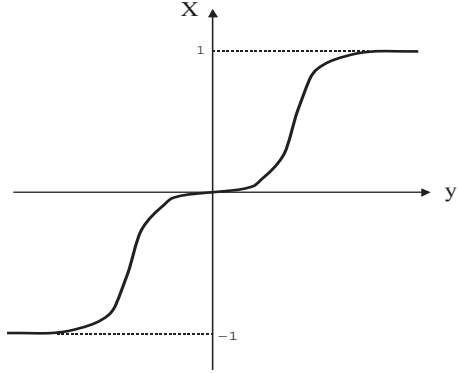}
\end{center}
\caption{Schematic picture of $X$ as a function 
of $y$.}
\label{X}
\end{figure}

\begin{figure}[htbp]
\begin{flushleft}
\leavevmode
\begin{eqnarray*}
\begin{array}{cc}
  \epsfxsize=5.5cm
  \epsfysize=5.5cm
  \epsfbox{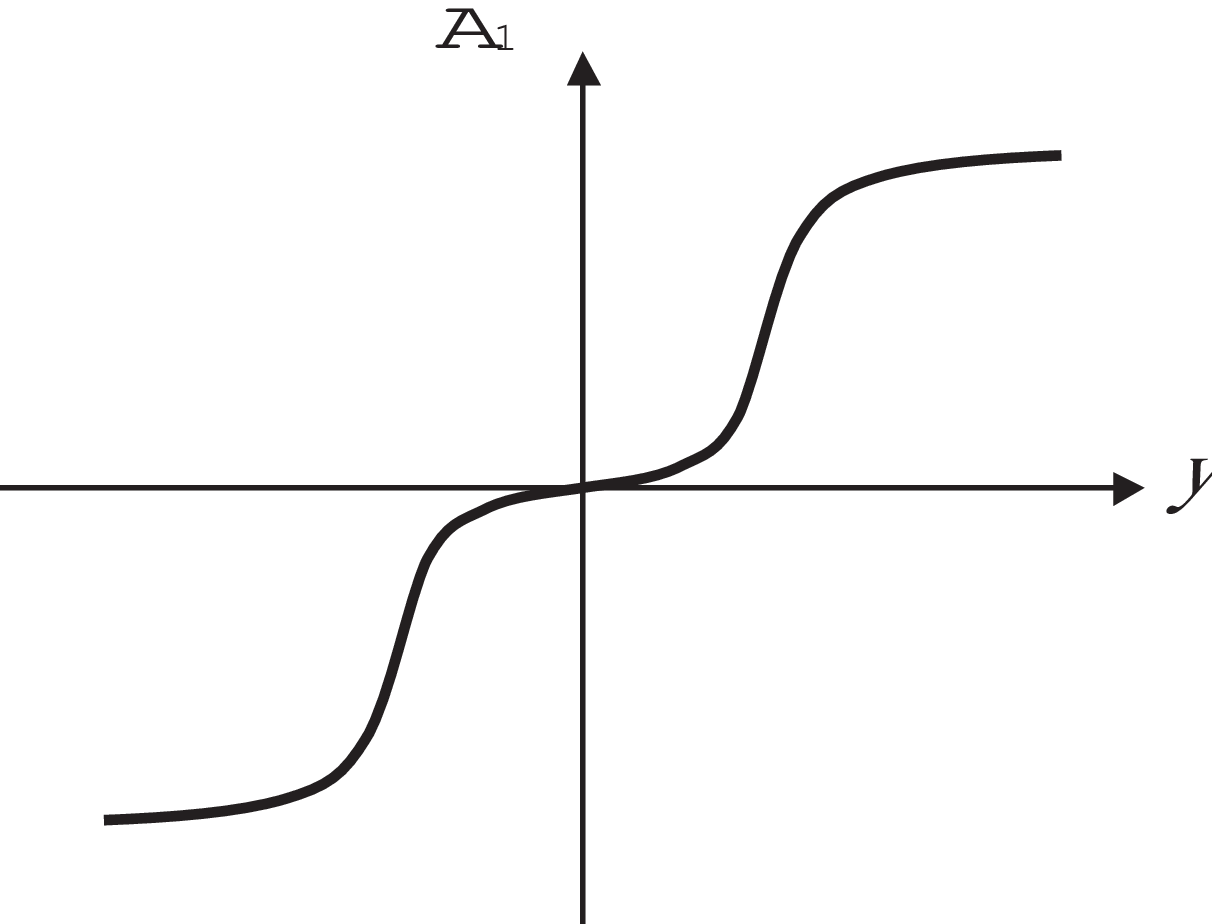} &
  \epsfxsize=5.5cm
  \epsfysize=5.5cm
  \hspace{1cm}
  \epsfbox{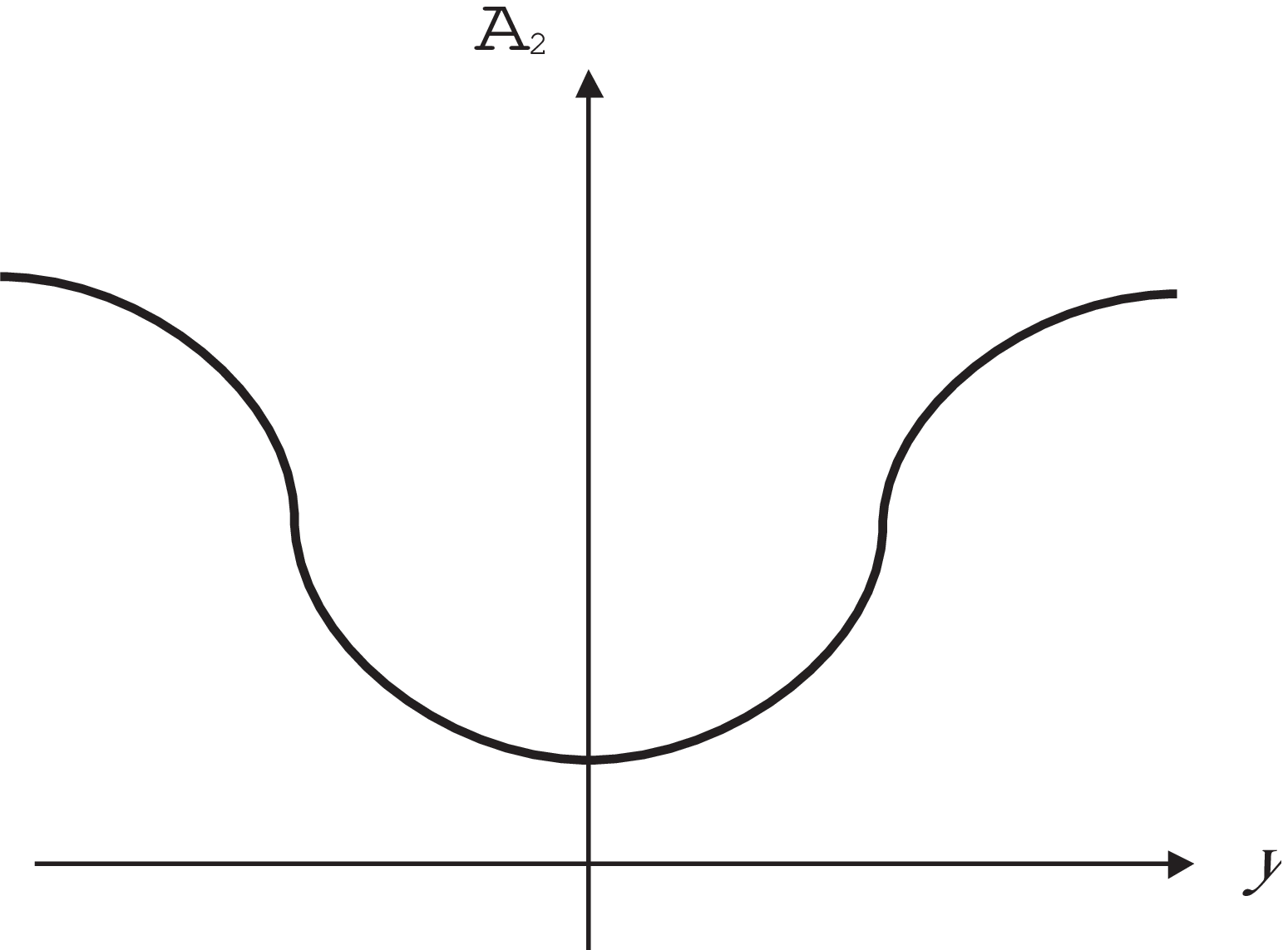} 
\\
  \mbox{(a) Scalar field $A_1$. } \quad  & \quad \quad 
\quad 
  \mbox{(b) Scalar field $A_2$. } 
\end{array} 
\end{eqnarray*} 
\caption{ Schematic pictures of the scalar fields as functions 
of $y$ : 
(a) $A_1$ and (b) $A_2$. 
}
\label{FIG:A}
\end{flushleft}
\end{figure}

In Figs.~\ref{X}, \ref{FIG:A} (a) and \ref{FIG:A} (b), 
we illustrate the behavior of the product $X$ and 
the charged scalar fields $A_1$ and $A_2$, respectively. 
We see that one of the charged scalar fields, $A_1$, vanishes 
at the point where the product $X$ vanishes. 
The other charged field, $A_2$, approaches 
very close to zero, although it does not vanish. 

The width of the wall can be identified as the length 
of the transition region in 
Eq.~(\ref{eq:transition-region}) 
\begin{equation}
\Delta y = \frac{\Delta \tilde{y}}{g \Lambda } = 
 \frac{1}
{4g\Lambda \sqrt{1 + \tilde{\kappa}^2}} 
{\rm log}\frac{4}{\tilde{\kappa}^2}. 
\end{equation}
This is of the same order as the inverse mass of the 
ground state of the vector field in 
Eq.~(\ref{eq:bound-mass-squared}). 
Therefore, 
we can make the mass of the lightest 
vector field as small as we like only at the cost of 
making the localization width larger. 

Lastly, let us note that the bosonic part of our model given 
in Eq.~(\ref{eq:boson_Lag}) can be 
promoted to a five-dimensional theory without 
further problems. 
Therefore, we can consider a five-dimensional version 
of our model of vector boson localization with this model. 
The only difference is that we can no longer make it a 
supersymmetric theory in five dimensions, because the 
theory in a system of dimension greater than or equal to five 
requires at least eight SUSY (${\cal N}=2$ SUSY theory). 
The ${\cal N}=2$ SUSY introduces more symmetry 
constraints that do not allow the potentials of our model.

\section*{Acknowledgements}
The authors thank Borut Bajc for useful communications. 
This work is supported in part by a Grant-in-Aid for Scientific 
Research from the Ministry of Education, Culture, Sports, 
Science and 
Technology, Japan No.13640269 (NS) and 
by the Special Postdoctoral Researchers Program at RIKEN (NM). 

\appendix
\section{Solving the BPS Equation} 
\label{ap:BPSeq}
The BPS equation in terms of the 
rescaled variables is given by
\begin{equation}
\frac{d \tilde{A}_1}{d \tilde{y}} = \sqrt{\tilde{A}_1^2 + 2 \tilde{\kappa}} 
\left[ 1 - \tilde{A}_1^2 (\tilde{A}_1^2 + 2 \tilde{\kappa}) \right].
\end{equation}
This is an integrable equation: 
\begin{eqnarray}
\tilde{y} &=& 
\int d \tilde{A}_1 
\frac{1}{\sqrt{\tilde{A}_1 + 2 \tilde{\kappa}}
[1-\tilde{A}_1^2 (\tilde{A}_1^2 + 2\tilde{\kappa})]} \\
&=& 
\int d \tilde{A}_1 
\frac{1}{\sqrt{\tilde{A}_1 + 2 \tilde{\kappa}}
[(a - \tilde{A}_1^2)(\tilde{A}_1^2 + b)]} \nonumber \\
&=& \int d \tilde{A}_1 
\frac{1}{\sqrt{\tilde{A}_1 + 2 \tilde{\kappa}}}
\left( \frac{1}{a - \tilde{A}_1^2} 
+ \frac{1}{\tilde{A}_1^2 + b} \right)
\frac{1}{a + b} \nonumber \\
&=& \int d \tilde{A}_1 
\frac{1}{\sqrt{\tilde{A}_1 + 2 \tilde{\kappa}}}
\left[ (\frac{1}{\sqrt{a} - \tilde{A}_1} 
+ \frac{1}{\sqrt{a} + \tilde{A}_1}) 
\frac{1}{2\sqrt{2}} \right. \nonumber \\
&& \left. + (\frac{1}{\tilde{A}_1 - i\sqrt{b}} 
- \frac{1}{\tilde{A}_1 + i\sqrt{b}}) 
\frac{1}{2i\sqrt{b}} \right]
\frac{1}{a + b}. \nonumber 
\end{eqnarray}
Then, using 
\begin{eqnarray}
&&\int \frac{d\tilde{A}}{\sqrt{\tilde{A}^2 + 2\kappa}} 
\frac{1}{\sqrt{a} + \tilde{A}} 
= \int \frac{d\tilde{A}'}
{\sqrt{(\tilde{A}'-\sqrt{a})^2 + 2\kappa}} 
\frac{1}{\tilde{A}'} \nonumber \\
&=& \int \frac{d\tilde{A}'}{\tilde{A}'} 
\frac{1}{\sqrt{\tilde{A}'{}^2 
- 2 \sqrt{a} \tilde{A}' + a + 2\kappa}} 
\nonumber \\
&=& \frac{1}{\sqrt{a + 2\kappa}} {\rm ln}
\left[ \frac{\tilde{A}'}
{-2\sqrt{a}\tilde{A}' + 2(a + 2\kappa) 
+ 2 
\sqrt{(a+2\kappa)(\tilde{A}'{}^2 - 2 \sqrt{a} 
\tilde{A}' + a + 2\kappa)}} \right] \nonumber \\
&=& \frac{1}{\sqrt{a + 2\kappa}} {\rm ln}
\left[ \frac{\tilde{A} + \sqrt{a}}{-2\sqrt{a}
(\tilde{A} + \sqrt{a}) + 2(a + 2\kappa) 
+ 2 \sqrt{(a + 2\kappa)(\tilde{A}^2 + 2\kappa)}} \right]
 \nonumber \\
&=& \frac{1}{\sqrt{a + 2\kappa}} {\rm ln}
\left[ \frac{\tilde{A} + \sqrt{a}}
{-2\sqrt{a}\tilde{A} + 4 \kappa 
+ 2 \sqrt{(a + 2\kappa)(\tilde{A}^2 + 2\kappa)}} \right]
\end{eqnarray}
and
\begin{equation}
\int \frac{d\tilde{A}}{\sqrt{\tilde{A}^2 + 2\kappa}} 
\frac{1}{\sqrt{a} - \tilde{A}} 
= - \frac{1}{\sqrt{a + 2\kappa}} 
{\rm ln}
\left[ \frac{-\tilde{A} + \sqrt{a}}
{2\sqrt{a}\tilde{A} + 4 \kappa 
+ 2 \sqrt{(a + 2\kappa)(\tilde{A}^2 + 2\kappa)}} 
\right], 
\end{equation}
we have
\begin{eqnarray}
&&\int \frac{d\tilde{A}}{\sqrt{\tilde{A}^2 + 2\kappa}} 
\left( \frac{1}{\sqrt{a} + \tilde{A}} 
+ \frac{1}{\sqrt{a} - \tilde{A}} \right)
\frac{1}{2\sqrt{a}} \nonumber \\
&&= \frac{1}{2\sqrt{a(a + 2\kappa)}} 
{\rm ln} \left( \frac{(\tilde{A} + \sqrt{a})
[\sqrt{a}\tilde{A} + 2\kappa 
+ \sqrt{(a + 2\kappa)(\tilde{A}^2 + 2\kappa)}]}
{(\sqrt{a} - \tilde{A})
[-\sqrt{a}\tilde{A} + 2\kappa 
+ \sqrt{(a + 2\kappa)
(\tilde{A}^2 + 2\kappa)}]} \right). 
\end{eqnarray}
Thus, we obtain the result 
(\ref{eq:BPSsol}). 

To obtain the asymptotic behavior of the BPS solution 
(\ref{eq:BPSsol}), we take the limit $y \to \infty$, 
finding  $\tilde{A} \to \sqrt{a}$ and 
\begin{eqnarray}
\tilde{y} - \tilde{y}_0 &\sim& 
\frac{1}{2(a+b) \sqrt{a(a + 2\tilde{\kappa})}} 
{\rm ln}\left( \frac{2\sqrt{a}(a + 2\tilde{\kappa} 
+ a + 2\tilde{\kappa})}
{(\sqrt{a} - \tilde{A})
(a + 2\tilde{\kappa} - a + 2\tilde{\kappa})} \right) 
+ {\rm const}  \nonumber \\
&\sim& -\frac{1}{2(a+b)\sqrt{a(a + 2 \tilde{\kappa})}} 
{\rm ln}(\sqrt{a} - \tilde{A}) 
+ {\rm const}  \nonumber \\
&\to& \sqrt{a} - \tilde{A} 
\sim e^{-2(a+b)\sqrt{a(a + 2 \tilde{\kappa})}
(\tilde{y} - \tilde{y}_0)}. 
\end{eqnarray}
 Also, for $y \to -\infty$, we obtain 
 $\tilde{A} \to -\sqrt{a}$ and 
\begin{eqnarray}
\tilde{y} - \tilde{y}_0 &\sim& 
\frac{1}{2(a+b) \sqrt{a(a + 2 \tilde{\kappa})}} 
{\rm ln}\left( \frac{(\tilde{A} + \sqrt{a})
(a + 2\tilde{\kappa} - a + 2\tilde{\kappa})}
{2\sqrt{a}(a + 2\tilde{\kappa} 
+ a + 2\tilde{\kappa})} \right) 
+ {\rm const}  \nonumber \\
&\sim& \frac{1}{2(a + b) \sqrt{a(a + 2 \tilde{\kappa})}} 
{\rm ln}(\sqrt{a} + \tilde{A}) 
+ {\rm const}  \nonumber \\
&\to& \sqrt{a} + \tilde{A} 
\sim e^{+2(a + b) \sqrt{a(a + 2 \tilde{\kappa})}
(\tilde{y} - \tilde{y}_0)}
.
\end{eqnarray}

\section{Mode Equation of the Vector Field} 
\label{ap:vector-mode}

To analyze the spectra of fields $v_m$, let us first show 
that there are no zero modes. 
If there is a zero mode, 
(\ref{eq:lin-EOM-v}) and (\ref{eq:lin-EOM-vy}) become 
\begin{eqnarray}
0 
&=& 
  v_\nu'' 
-\partial_\nu \partial^\mu v_\mu 
-\partial_\nu v_y' 
-V(y)v_\nu ,
\label{eq:zeromode1}
\\
0 
&=& - \partial^\mu v_\mu' 
-V(y)v_y
, 
\label{eq:zeromode2}
\end{eqnarray}
because a zero mode is defined by $\square v_m=0$. 
Then, applying $\partial^\nu$ to (\ref{eq:zeromode1}), 
we obtain 
\begin{eqnarray}
0 
&=& 
(\partial_y^2 -V(y)) \partial^\nu v_\nu.
\label{eq:zeromode3}
\end{eqnarray}
The positive definite potential $V(y)$ allows 
no normalizable solution, that is no solution satisfying 
$
 \partial^\nu v_\nu 
=
0
$. 
Then, inserting this result into (\ref{eq:zeromode2}), 
we obtain 
immediately 
\begin{eqnarray}
 v_y 
&
=
& 
0. 
\label{eq:zeromode6}
\end{eqnarray}
Then, inserting these results into (\ref{eq:zeromode1}), 
we obtain 
$
0 
=
(\partial_y^2 -V(y)) v_\nu 
,
$ implying 
\begin{eqnarray}
 v_\nu = 0 . 
\label{eq:zeromode8}
\end{eqnarray}
Therefore we find that there is no zero mode. 

Next, we decompose the linearized equations of motion 
into transverse and longitudinal components. 
Since there is no zero mode, we can separate the 
transverse component $\tilde v_\mu$ from the longitudinal 
one $\partial^\lambda v_\lambda$, 
as in Eq.~(\ref{eq:transverse-comp}). 
Equation (\ref{eq:lin-EOM-v}) can be rewritten in terms of the 
transverse and longitudinal components as 
\begin{eqnarray}
0 
&=& \square \tilde v_\nu 
+  \tilde  v_\nu'' 
-V(y) \tilde v_\nu 
+ \partial_\nu\left[{1 \over \square} 
\left(\partial_y^2 -V(y) \right) \partial^\lambda v_\lambda 
- v_y' \right] .
\label{eq:lin-EOM-trans}
\end{eqnarray}
By applying $\partial^\nu$ to this, we can eliminate the 
transverse component and obtain 
\begin{eqnarray}
0 
&=& 
\left(\partial_y^2 -V(y) \right) \partial^\lambda v_\lambda 
- \square v_y' . 
\label{eq:lin-EOM-long}
\end{eqnarray}
Inserting this result into (\ref{eq:lin-EOM-trans}), 
we obtain the linearized equations of motion for 
the transverse component, as given in Eq.~(\ref{eq:lin-EOM-trans2}). 
Thus we find that the linearized equations of motion 
for the transverse component are decoupled from the 
longitudinal component $\partial^\lambda v_\lambda$ 
and the $y$ component $v_y$. 

The linearized equations of motion for the 
longitudinal component can also be obtained. 
By applying $\partial_y$ to Eq.~(\ref{eq:lin-EOM-vy}), 
we obtain 
\begin{eqnarray}
0 
&=& \square v_y' 
- \partial^\lambda v_\lambda'' 
-\left(V(y) v_y \right)' 
.
\label{eq:lin-EOM-vy2}
\end{eqnarray}
Then, adding this to (\ref{eq:lin-EOM-long}), 
we obtain the longitudinal component 
in terms of $v_y$: 
\begin{eqnarray}
\partial^\lambda v_\lambda 
&=& 
-{1 \over V(y)}\left(V(y) v_y \right)' 
.
\label{eq:long-vy}
\end{eqnarray}
Inserting this relation into Eq.~(\ref{eq:lin-EOM-vy}), 
we finally obtain the linearized 
equations of motion for $v_y$: 
\begin{eqnarray}
0 
&=& \square v_y 
+ \left({1 \over V}(Vv_y)'\right)' 
-V(y) v_y 
.
\label{eq:lin-EOM-vy3}
\end{eqnarray}

\section{Variational Approach for the Ground State} 
\label{ap:variation}

Employing the ground state wave function for the rigid wall 
potential (\ref{eq:rigid-wall}) 
with width $2a/\mu$, we obtain 
the expectation value of the Hamiltonian in 
Eq.~(\ref{eq:vec-mode-eq}) to be 
\begin{eqnarray}
\langle H \rangle (a) = \mu^2 
\left[
\left({\pi \over 2a}\right)^2 + 
{\alpha^2 \over \pi} {1 \over {2a \over \alpha \pi} 
+ {\alpha \pi \over 2a}} {\rm sinh} {2a \over \alpha}
\right].
\end{eqnarray}
The minimum of this expectation value 
is realized at stationary point with respect to $a$, 
which is accurately approximated by 
\begin{eqnarray}
0={1 \over \mu^2} {d \langle H \rangle \over da} \approx 
-\left({\pi \over 2}\right)^2 {2 \over a^3}
+ {\alpha^2 \over 2a} {\rm e}^{2a \over \alpha} 
\end{eqnarray}
for large values of $a/\alpha$, 
because we are interested in the case described by 
Eq.~(\ref{eq:alpha}). 
This transcendental equation can be solved iteratively as 
\begin{eqnarray}
{a \over \alpha}= 
{\rm log}\left({\pi \over \alpha^2}\right) 
- 
{\rm log}\left({a \over \alpha}\right) 
\approx 
{\rm log}\left({\pi \over \alpha^2}\right) 
- 
{\rm log}{\rm log}\left({\pi \over \alpha^2}\right) 
+ \cdots, 
\end{eqnarray}
determining the width 
 $a$ as given in Eq.~(\ref{eq:width}). 
We see that this result confirms the validity of our approximation for 
large $a/\alpha$. 
At the minimum of $\langle H \rangle$ with respect to 
the variational parameter $a$ for small $\alpha$, 
we find that the kinetic energy is dominant, and that 
the upper bound 
for the ground state mass squared 
is given by (\ref{eq:best-bound}).

\end{document}